\documentstyle[prd,tighten,aps]{revtex}
\begin{document}
\draft

\twocolumn[\hsize\textwidth\columnwidth\hsize\csname
@twocolumnfalse\endcsname
\title{\bf Ringholes and closed timelike curves}

\author{Pedro F. Gonz\'alez-D\'{\i}az}
\address{Centro de F\'{\i}sica ``Miguel Catal\'an'',
Instituto de Matem\'aticas y F\'{\i}sica Fundamental,\\
Consejo Superior de Investigaciones Cient\'{\i}ficas,
Serrano 121, 28006 Madrid (SPAIN)}
\date{March 1, 1996}

\maketitle

\begin{abstract}
It is shown that in a classical spacetime with multiply connected
space slices having the topology of a torus, closed timelikes
curves are also formed. We call these spacetimes ringholes.
Two regions on the torus surface can be
distinguished which are separated by angular horizons. On one
of such regions (that which surrounds the maximum circunference
of the torus) everything happens like in spherical wormholes,
but the other region (the rest of the torus surface), while still
possessing a chronology horizon and non-chronal region, behaves
like a converging, rather than diverging, lens and corresponds to
an energy density which is always positive for large speeds at
or near the throat. It is speculated
that a ringhole could be converted into a time machine to
perform time travels by an observer who would never encounter
any matter that violates the classical averaged weak energy condition.
Based on a calculation of vacuum fluctuations, it is also seen
that the angular horizons can prevent the emergence of
quantum instabilities near the throat.

\end{abstract}
\pacs{PACS number(s): 04.20.Cv, 04.62.+v }

\vskip2pc]

\renewcommand{\theequation}{\arabic{section}.\arabic{equation}}

\section{\bf Introduction}

Time travels have become a scientific possibility in recent
years [1]. Spacetimes possessing closed timelike curves (CTCs)
were originally discussed by G"del [2], but it only was after
recognition that negative-energy states can occur in semiclassical
general relativity that time travels have been considered in greater
detail. Recently, the subject of CTC was revived in two apparently
quite different contexts. The first involves tunneling through
wormholes in four-dimensional Einstein gravity [3]. The second,
which we shall not discuss here, concerns solutions to gravity
with infinite parallel cosmic strings which move with high
velocity relative to each other [4]. Although the latter does
not entail negative energies, the moving straight strings can only
produce CTCs if they are infinite in length [5] and could
therefore not exist in a finite universe.

Time travels through four-dimensional wormholes entail an inward
pressure on the tunnel so great that it ultimately meant a
matter with negative energy density [3]. Although negative
energy is becoming commonplace in quantum and semiclassical
gravity [6], it would imply violation of the averaged weak energy
condition in classical general relativity. This situation
is deeply troublesome and, if one calculate quantum fluctuations
of vacuum in wormholes where one can provide them with a more natural
context for negative energies to occur, then divergences of
the stress-energy tensor arise [7] that may make the problem
even more acute [5]. It appears therefore that for a classical
solution to Einstein equations with CTCs to offer an acceptable
scenario for time travels, such a solution must provide with
finite geometrical objects able to live in our universe and
simultaneously entail classical stress-energy tensors with
positive energy density.

The ultimate reason for traversible wormholes to have negative
energy densities is that their very geometry necessarily
requires any bundle of rays entering one wormhole's mouth
radially converging, to emerge from the other mouth radially
diverging [8]. This diverging-lens action, on the other hand,
prevents the spacetime structure from pinching off at the
throat and, moreover, is responsible for the existence of
CTCs in it. However, though the diverging-lens effect protects
the wormhole against the classical instability produced by
the wave's pileup at or near the throat, it is unable to
prevent the wormhole from quantum instabilities originated
by electromagnetic vacuum fluctuations [7,8].
This situation has led Hawking to suggest his
chronology protection conjecture, according to which the laws
of physics do not allow the appearance of CTCs [5].

A possible way out from this situation could be making recourse
to tunnels with topologies other than spherical. The most
obvious candidate for one such topologies is the torus. One
would then tunnel from a region of Euclidean space by means of
a torus whose transverse cross-sectional area would
shrink down to a minimum size at the throat,
increasing thereafter to finally be embedded again in
another region of Euclidean space (Fig. 2). We shall denote
this geometric construct a ringhole. The interesting thing one
would expect about these ringholes is that, although bundles
of rays originally converging in the direction tangent to the
torus at points on the maximum circunference will still emerge
diverging in the same direction, bundles of rays originally
diverging in the direction tangent to the torus at points on
the minimum circunference would now emerge converging in the
same direction at the other ringhole's mouth. Then, if the
surface surrounding the maximum circunference is able to
prevent the whole geometrical construct from pinching off at
the throat, by traversing the ringhole along a world line
which would pass through
the neighborhood of the minimum circunference, one might
expect the appearance of CTCs without any violation of the averaged
weak energy condition [9], whenever one ringhole's mouth is kept
at rest and the other moves at sufficiently high velocity.

In this paper we shall explore this possibility by constructing
the metric of a traversible ringhole which is solution to
the Einstein equations, and converting it into a time machine.
It is seen that also in this case, on the neighborhood of the
maximum circunference of the torus only the leftward chronology
(Cauchy) horizon and non-chronal region with CTCs survive,
while the averaged weak energy condition is violated at or near
the throat. However, on the neighborhood of the minimum circunference
of the torus, what survive are the rightward chronology (Cauchy)
horizon and non-chronal region with CTCs, while, quite remarkably,
the averaged weak energy condition holds at or near the
throat. Thus, we find no classical obstruction for an advanced
civilization to construct a time machine with toroidal topology
by using traversible ringholes with just one of their mouths
kept moving with respect to the other mouth, whenever the travel
itinerary passes through the throat along a world line near the
minimum toroidal circunference. By using the method of Kim and
Throne [7], it has been also shown that, in this case,
vacuum fluctuations
do not destabilize the accelerating ringhole either.

The body of this paper is organized as follows: In Sec. II we
review the way in which the static metric describing the spacetime
generated by a distribution of matter with the topology of a torus
can be obtained, and discuss the existence of apparent angular
horizons of this metric. In Sec. III we consider the existence
of what we call ringholes, that is tunnels in Lorentzian spacetime
with toroidal symmetry, deriving their metric so as the
characteristics of the stress-energy tensor that make these
ringholes compatible with general relativity. Lensing actions in
the ringhole space are also considered. Then in Sec. IV we convert
ringhole into time machine and analyse the causal and non-causal
structure of the resulting space in relation with the possibility
of using accelerating ringholes to travel back- and forward in
time, and in Sec. V we compute the quantum effect of vacuum
polarization following the procedure of Kim and Thorne in [7].
Finally, we summarize and briefly comment on our results in Sec. VI.

\section{\bf Static metric on the torus}
\setcounter{equation}{0}

Let us consider the gravitational field
created by a distribution of matter having the symmetry of a torus.
We shall obtain the static space-time metric corresponding to any
constant 2+1 surface with such a symmetry.

Defining the Cartesian coordinates as in Fig.1, it follows
$x=m\sin\varphi_{1}$, $y=m\cos\varphi_{1}$, $z=b\sin\varphi_{2}$,
so that
\begin{equation}
dx^{2}+dy^{2}+dz^{2}=m^{2}d\varphi_{1}^{2}+b^{2}d\varphi_{2}^{2},
\end{equation}
with
\begin{equation}
m=a-b\cos\varphi_{2}
\end{equation}
\begin{equation}
n=b-a\cos\varphi_{2},
\end{equation}
$a$ and $b$ being the radius of the circunference generated by the
circular axis of the torus and that of a torus section, respectively.

We can assume then for the static metric corresponding to a distribution
of matter with the symmetry of a torus the general expression
\begin{equation}
ds^{2}=-e^{\Phi}dt^{2}+e^{\Psi}dr^{2}+m^{2}d\varphi_{1}^{2}+
b^{2}d\varphi_{2}^{2},
\end{equation}
where
\begin{equation}
r=\left(a^{2}+b^{2}-2ab\cos\varphi_{2}\right)^{\frac{1}{2}},
\end{equation}
and $\Phi$ and $\Psi$ will generally depend on $r$ and $t$.

Denoting by $x^{0}, x^{1}, x^{2}, x^{3}$, respectively, the coordinates
on the torus $ct, r, \varphi_{1}$ and $\varphi_{2}$, the nonzero
components of the metric tensor take the general form
$g_{00}=-e^{\Phi}$, $g_{11}=e^{\Psi}$, $g_{22}=m^{2}$ and
$g_{33}=b^{2}$. Using then
\begin{equation}
\frac{dm}{dr}=\frac{r}{a},\;\; \frac{dm}{d\varphi_{2}}=b\sin\varphi_{2},
\;\; \frac{dn}{dr}=\frac{r}{b},\;\; \frac{dn}{d\varphi_{2}}=a\sin\varphi_{2}
\end{equation}
\begin{equation}
\frac{da}{dr}=\frac{r}{m},\;\;\;\; \frac{db}{dr}=\frac{r}{n},
\end{equation}
one can calculate the components of the affine connection and hence
those of the Ricci curvature tensor. Finally, one has for the
Einstein equations corresponding to the gravitational field of a
matter distribution with the symmetry of a torus
\[e^{-\Psi}\left[r^{2}\left(\frac{1}{m^{2}a^{2}}+\frac{1}{n^{2}b^{2}}\right)
+\left(\frac{1}{ma}+\frac{1}{nb}\right)\left(\frac{1}{2}r\Psi '-1\right)\right.\]
\begin{equation}
\left.-\frac{r^{2}}{mnab}\right]
-\frac{\cos\varphi_{2}}{mb}
=\frac{8\pi G}{c^{4}}T_{0}^{0}
\end{equation}
\vspace{.5cm}
\[-e^{-\Psi}\left[\frac{1}{2}\left(\frac{1}{ma}+\frac{1}{nb}\right)\Phi 'r
+\frac{r^{2}}{mnab}\right]\]
\begin{equation}
-\frac{\cos\varphi_{2}}{mb}=\frac{8\pi G}{c^{4}}T_{1}^{1}
\end{equation}
\vspace{.5cm}
\[e^{-\Psi}\left[\frac{r^{2}}{n^{2}b^{2}}-\frac{r(\Phi '-\Psi ')}{2nb}-\frac{1}{nb}\right]\]
\begin{equation}
+F(r,\varphi_{2},\Psi,\Phi)=\frac{16\pi G}{c^{4}}T_{2}^{2}
\end{equation}
\vspace{.5cm}
\[e^{-\Psi}\left[\frac{r^{2}}{m^{2}a^{2}}-\frac{r(\Phi '-\Psi ')}{2ma}-\frac{1}{ma}\right]\]
\begin{equation}
+F(r,\varphi_{2},\Psi,\Phi)=\frac{16\pi G}{c^{4}}T_{3}^{3}
\end{equation}
\vspace{.5cm}
\begin{equation}
-\frac{1}{2}r\dot{\Psi}\left(\frac{1}{ma}+\frac{1}{nb}\right)e^{-\Psi}
=\frac{8\pi G}{c^{4}}T_{0}^{1},
\end{equation}
where the prime denotes differentiation with respect to $r$,
the overhead dot accounts for time derivative, and
\[F(r,\varphi_{2},\Psi,\Phi)
=-\frac{1}{2}e^{-\Psi}\left(\Phi ''
-\frac{1}{2}\Phi '\Psi '+\frac{1}{2}\Phi '^{2}\right)\]
\begin{equation}
+\frac{1}{2}e^{-\Phi}\left(\ddot{\Psi}-\frac{1}{2}\dot{\Psi}\dot{\Phi}
+\frac{1}{2}\dot{\Psi}^{2}\right).
\end{equation}
General solutions to these equations look quite complicate. Nevertheless,
we shall derive a solution for the vacuum case
whenever $a$ and $b$ are constant.

Outside the space region occupied by the matter which creates the
gravitational field we must have $T_{i}^{i}=0$, $i=0,1,...3$. For
this vacuum case, using $r^{2}=ma+nb$,
one can obtain from the first pair of Einstein
equations
\begin{equation}
\frac{1}{2}r\left(\Phi '+\Psi '\right)=1-\left(\frac{ma}{nb}+\frac{nb}{ma}\right).
\end{equation}
Likewise, from the second pair of Einstein equations in the vacuum we arrive at
\begin{equation}
\frac{1}{2}r\left(\Phi '-\Psi '\right)=1+\left(\frac{ma}{nb}+\frac{nb}{ma}\right).
\end{equation}
Clearly, in the vacuum we also have from (2.12)
\begin{equation}
\dot{\Psi}=0.
\end{equation}

If we assume then a torus surface where the radii $a$ and $b$ are
both constant (which we shall denote a constant torus),
the solutions to (2.14) and (2.15) become
\begin{equation}
\Phi=\ln\frac{r^{2}}{C_{2}}
\end{equation}
\begin{equation}
\Psi=\ln\left[\frac{C_{1}}{r^{4}\left(1-\frac{A^{2}}{r^{4}}\right)^{2}}\right],
\end{equation}
where $A=a^{2}-b^{2}$ and $C_{1}$ and $C_{2}$ are integration
constants which
would be given in terms of the mass distribution with the symmetry
of a torus. The expression of these constants will not be of any
importance in the discussion to follow.

Thus, the following three-dimensional metric is finally obtained
for a torus symmetry with constant radii $a$ and $b$:
\[ds^{2}=-C_{2}r^{2}dt^{2}+\]
\[\left[\frac{C_{1}}{r^{4}\left(1-\frac{A^{2}}{r^{4}}\right)^{2}}
+\frac{4b^{2}}{2B-r^{2}\left(1+\frac{A^{2}}{r^{4}}\right)}\right]dr^{2}
+m^{2}d\varphi_{1}^{2}\]
\begin{equation}
=-C_{2}r^{2}dt^{2}
+b^{2}\left[1+\frac{C_{1}a^{2}\sin^{2}\varphi_{2}}{r^{6}\left(1-
\frac{A^{2}}{r^{4}}\right)^{2}}\right]d\varphi_{2}^{2}
+m^{2}d\varphi_{1}^{2},
\end{equation}
where $B=a^{2}+b^{2}$, with $a$ and $b$ constant and $a>b$.
Metric (2.19) is defined for $0\leq t\leq\infty$, $a-b\leq r\leq a+b$,
$0\leq\varphi_{1},\varphi_{2}\leq 2\pi$, and describes the
space-time geometry on a torus generated by varying angles
$\varphi_{1}$ and $\varphi_{2}$ only; $i.e.$: the space-time
geometry on a torus with constant radii $a$ and $b$ generated
by moving about its entire surface.

Of most interest is the variation of the metric with angle
$\varphi_{2}$. On $\varphi_{2}=\pi$, where $r=r_{max}=a+b$
(the radius of the maximum circunference of the torus) the
metric becomes singular. As $\varphi_{2}$ decreases from $\pi$
to $\varphi_{2}^{c}=\arccos\frac{b}{a}$, $r$ decreases from
$r_{max}$ to $r=\sqrt{A}$ where the metric becomes singular as
well. The metric is again singular on $\varphi_{2}=0$ where
$r=r_{min}=a-b$ (the radius of the minimum circunference of
the torus). Of course, the same singularity as that on $\varphi_{2}^{c}$
will also occur on $2\pi-\varphi_{2}^{c}$. One would then cut the
points on the two extremum circunferences of the torus and
on the two circunferences for which $r=\sqrt{A}$ out of the
manifold defined by the coordinates $t,r,\varphi_{1}$ (or
$t,\varphi_{1},\varphi_{2}$), and so divide this manifold
into four disconnected components.

However, one would expect
the singularities on $r=r_{min}$ and $r=r_{max}$ to play an
analogous role to that of the trivial singularities of polar
coordinates when $\theta=0$ and $\theta=\pi$. Moreover,
calculation of the curvature invariants, such as $R^{ijkl}R_{ijkl}$,
shows that all of these invariants are finite on the above
points where the metric becomes singular in coordinates
$t,r,\varphi_{1}$ (or $t,\varphi_{1},\varphi_{2}$). This
strongly suggests that all of these singularities are not
true singularities but only arise from the choice of
coordinates [9]. In fact, if we first introduce the new real
coordinates
\begin{equation}
u=t+\frac{C_{3}}{4A}\ln\left(\frac{A-r^{2}}{A+r^{2}}\right),\;\;\;
v=t-\frac{C_{3}}{4A}\ln\left(\frac{A-r^{2}}{A+r^{2}}\right),
\end{equation}
with $C_{3}=\sqrt{\frac{C_{1}}{C_{2}}}$, the metric takes the
form
\begin{equation}
ds^{2}=-C_{2}r^{2}dudv
+\frac{4b^{2}dr^{2}}{2B-r^{2}\left(1+\frac{A^{2}}{r^{4}}\right)}
+m^{2}d\varphi_{1}^{2},
\end{equation}
where
\[r^{2}=A\tanh\left[\frac{A(u-v)}{C_{3}}\right],\]
which still is singular on $r=r_{max}$ and $r=r_{min}$.
Introducing then the additional complex coordinate
\begin{equation}
w=i\varphi_{1}+\frac{b}{2\sqrt{A}}\arcsin\left[\frac{a\left(r^{2}-
A\right)}{b\left(r^{2}+A\right)}\right],
\end{equation}
allows one to re-write metric (2.19) into the final form in
coordinates $u,v,w$
\begin{equation}
ds^{2}=-C_{2}r^{2}dudv
+m^{2}\left|dw\right|^{2},
\end{equation}
where
\[i\varphi_{1}=\frac{1}{2}\left(w-w^{*}\right)\]
\[r^{2}=A\left(\frac{1+\sin\varphi_{2}^{c}\sin x}{1-
\sin\varphi_{2}^{c}\sin x}\right),\]
in which $x=\frac{\sqrt{A}}{b}\left(w+w^{*}\right)$ is real and
$^{*}$ denotes complex conjugation.

It is easy to see that metric (2.23) is real and in fact regular
everywhere in the complete three-manifold for which
$r_{min}\leq r\leq r_{max}$.
We note finally that by rotating $\varphi_{1}\rightarrow\tilde{\varphi}_{1}
=-i\varphi_{1}$, coordinate $w$ unfolds into two
distinct real coordinates
$w\rightarrow w_{1}$ and $w^{*}\rightarrow w_{2}$, and hence metric (2.23)
becomes
\[ds^{2}\rightarrow d\tilde{s}^{2}=
ds^{2}-m^{2}\left(\left|dw\right|^{2}-dw_{1}dw_{2}\right).\]

\section{\bf Ringhole spacetime}
\setcounter{equation}{0}

One would expect that creation of traversible ringholes respecting
Einstein equations in such a way that classical general relativity
be valid everywhere, must be accompanied by CTCs in some non-chronal
spacetime sectors that would appear at late time, and by violation of the
averaged weak energy condition only on some restricted, classically forbidden
spatial regions.

A static ringhole having toroidal topology would be traversible if
a two-torus surrounding one of its mouths where spacetime is nearly flat
can be regarded as an outer trapped surface to an observer looking
through the ringhole from the other mouth [3]. The static
spacetime metric for one such single, traversible ringholes can
be written in the form
\begin{equation}
ds^{2}=-dt^{2}+\left(\frac{n_{l}}{r_{l}}\right)^{2}dl^{2}
+m_{l}^{2}d\varphi_{1}^{2}+\left(l^{2}+b_{0}^{2}\right)d\varphi_{2}^{2},
\end{equation}
where $-\infty < t <+\infty$, $-\infty < l < +\infty$, $\varphi_{1}$
and $\varphi_{2}$ are as given for the torus metric (2.19), $l$ is
the proper radial distance of each transversal section of the
torus, and
\begin{equation}
m_{l}=a-\left(l^{2}+b_{0}^{2}\right)^{\frac{1}{2}}\cos\varphi_{2}
\end{equation}
\begin{equation}
n_{l}=\left(l^{2}+b_{0}^{2}\right)^{\frac{1}{2}}-a\cos\varphi_{2}
\end{equation}
\begin{equation}
r_{l}=\left[a^{2}+l^{2}+b_{0}^{2}-2\left(l^{2}+
b_{0}^{2}\right)^{\frac{1}{2}}a\cos\varphi_{2}\right]^{\frac{1}{2}}
\end{equation}

Metric (3.1) is of special interest as it gives us a particularly
simple example of a traversible ringhole which is readily
generalizable. One can convert (3.1) into the more
general static ringhole metric (note that $rdr=ndb$)
\[ds^{2}=-dt^{2}e^{2\Phi}+\frac{dr^{2}}{1-\frac{K(b)}{b}}
+m^{2}d\varphi_{1}^{2}+b^{2}d\varphi_{2}^{2}\]
\begin{equation}
=-dt^{2}e^{2\Phi}+\left(\frac{n}{r}\right)^{2}dl^{2}
+m^{2}d\varphi_{1}^{2}+b^{2}d\varphi_{2}^{2}
\end{equation}
if we let $l=\pm\left(b^{2}-b_{0}^{2}\right)^{\frac{1}{2}}$,
$\Phi=0$ and $K(b)=\frac{b_{0}^{2}}{b}$, where the minus sign
applies on the left side of the throat and the plus sign does on
the right side. $\Phi$ will generally be now given as a function of
the ringhole mass and the geometric parameters $a$ and $b$. At
the throat $l=0$, so that the throat radius becomes $b_{0}$.

Metric (3.5) can be regarded as a generalization to toroidal symmetry
from the static, spherical wormhole metric [3]. The reduction
transformations
$a\rightarrow 0$, $\varphi_{2}\rightarrow\theta+\frac{\pi}{2}$,
$\varphi_{1}\rightarrow\phi$, where $\theta$ and $\phi$ are the
angular polar coordinates on the two-sphere, leads to $r=n=b$,
$m=r\sin\theta$, with which metric (3.5) reduces to
\begin{equation}
ds^{2}=-dt^{2}e^{2\Phi}+dl^{2}
+r^{2}\left(d\theta^{2}+\sin^{2}\theta d\phi^{2}\right),
\end{equation}
which, in fact, is the static metric used by Morris et al. [3] if
we take $r\cong |l|-M\ln\left(\frac{|l|}{b_{0}}\right)$ and
$\Phi\cong -\frac{M}{|l|}$ far from the throat, and $M$ is regarded
to be the wormhole mass. One must add $\frac{\pi}{2}$ to $\theta$
to get $\varphi_{2}$ because of the different choice of the ringhole
symmetry axis which is here taken to be the axis $z$ of Fig.1.

Metric (3.5) will be viewed to represent tunneling through a
traversible ringhole, and satisfy Einstein equations for
convenient stress-energy tensors for the following reasons:

\vspace{.5cm}

\noindent (i) As $l$ increases from $-\infty$ to 0, $b$ decreases
monotonously from $+\infty$ to a minimum value $b_{0}$ (the throat
radius), and as $l$ increases onward to $+\infty$, $b$ increases
monotonously to $+\infty$ again.

\vspace{.5cm}

\noindent (ii) For (3.5) to describe a ringhole we must embed it
in a three-dimensional Euclidean space at a fixed time $t$ [3].
We should consider a three-geometry that would respect the symmetry of
a torus and satisfy $a\geq b<l$, so it will suffice confining
attention to the maximum- and
minimum-circunference slices, i.e. $\varphi_{2}=\pi, 0$,
through it. In the first case, $r=m=n=a+b$, and
\begin{equation}
ds^{2}=\frac{dr^{2}}{1-\frac{b_{0}^{2}}{b^{2}}}+r^{2}d\varphi_{1}^{2}.
\end{equation}
We wish now visualize the slice (3.7) as removed from spacetime
(3.5) and embedded in three-dimensional Euclidean space. For the
embedding space, one takes a space with cylindrical coordinates
$z,r,\phi$,
\begin{equation}
ds^{2}=dz^{2}+dr^{2}+r^{2}d\phi^{2}.
\end{equation}
The embedded surface will be axially symmetric and can therefore
be described by a function $z=z(r)$. From (3.8) we have
\begin{equation}
ds^{2}=\left[1+\left(\frac{dz}{dr}\right)^{2}\right]dr^{2}+r^{2}d\phi^{2}.
\end{equation}
Metric (3.9) will be the same as metric (3.7) if we identify the
coordinates $r,\phi$ of the embedding space with the coordinates
$r,\varphi_{1}$ of the ringhole spacetime, and if we require the
function $z(r)$ to satisfy
\begin{equation}
\frac{dz}{dr}=\left(\frac{b^{2}}{b_{0}^{2}}-1\right)^{-\frac{1}{2}}.
\end{equation}
This equation displays the way in which, provided we have a fixed
$a\geq b$, the function $b\equiv b(l)$ shapes the ringhole spatial
geometry. In Fig. 2 we picture the surface $z(r)$.

If we confine attention to the minimum-circunference slice,
$\varphi_{2}=0$, through the geometry of three-space at a fixed
time, then $r=m=-n=a-b$ and the metric (3.7) is obtained again.
Thus, no matter the election of slice, we always achieve the
same condition (3.10).

\vspace{.5cm}

\noindent (iii) From the Einstein equations (2.8) and (2.9) we
obtain for the metric components of (3.5) with $\Phi=0$:
\[-\frac{b_{0}^{2}}{nb^{3}}\left(2+\frac{ma}{nb}+\frac{nb}{ma}\right)+\]
\begin{equation}
\left(1-\frac{b_{0}^{2}}{b^{2}}\right)\left(\frac{ma}{n^{2}b^{2}}
+\frac{nb}{m^{2}a^{2}}\right)=\frac{8\pi G}{c^{4}}\left(T_{0}^{0}-T_{1}^{1}\right).
\end{equation}
The stress-energy tensor components $T_{0}^{0}$ and $T_{1}^{1}$
in (3.11) cannot be directly given in terms of, respectively,
an energy density, $\rho$, and a tension per unit area, $\sigma$,
for the toroidal symmetry, because these tensor components should
explicitly depend on $r$ (see Fig. 1) whereas $\rho$ and $\sigma$
must be defined in function of the normal to a surface element on
the torus, along the $b$-direction. Since $\frac{dr}{db}=\frac{n}{r}$,
in the neighborhood of the ringhole throat ($b\simeq b_{0}$), we have
then
\[\rho c^{2}-\sigma=\left(\frac{n}{r}\right)^{3}\left(T_{0}^{0}-T_{1}^{1}\right)\simeq\]
\begin{equation}
-\frac{c^{4}n^{2}b_{0}^{2}}{8\pi Gr^{3}b^{3}}\left(2+\frac{ma}{nb}
+\frac{nb}{ma}\right).
\end{equation}

The requirement that ringholes be connectible to asymptotically
flat spacetime entails at the throat that the embedding surface
flares outward for $2\pi-\varphi_{2}^{c}>\varphi_{2}>\varphi_{2}^{c}$
and "flares inward" for $-\varphi_{2}^{c}<\varphi_{2}<\varphi_{2}^{c}$.
One would expect these flarings to be mathematically expressed by
the conditions
\begin{equation}
\frac{d^{2}r}{dz^{2}}>0 \;\;\; for\; 2\pi-\varphi_{2}^{c}>\varphi_{2}>\varphi_{2}^{c}
\end{equation}
\begin{equation}
\frac{d^{2}r}{dz^{2}}<0 \;\;\; for\; -\varphi_{2}^{c}>\varphi_{2}>\varphi_{2}^{c},
\end{equation}
at or near the throat.

From (3.10) we can obtain
\begin{equation}
\frac{d^{2}r}{dz^{2}}=\frac{br}{nb_{0}^{2}},
\end{equation}
which in fact satisfies both conditions (3.13) and (3.14). Now, since
$r^{3}b^{3}$ and $b_{0}^{2}$ are both positive, from (3.12) we obtain
\[sign\left[\rho c^{2}-\sigma\right]=sign\left[-\left(2+\frac{ma}{nb}+\frac{nb}{ma}\right)\right]\]
at or near the ringhole throat. It is easy to check that, since $r$ and $m$
are always positive for any value of $\varphi_{2}$, and $n$ is positive
for $2\pi-\varphi_{2}^{c}>\varphi_{2}>\varphi_{2}^{c}$ and negative
for $-\varphi_{2}^{c}<\varphi_{2}<\varphi_{2}^{c}$, we have finally
\begin{equation}
\rho c^{2}-\sigma <0 \;\;\; for\; 2\pi-\varphi_{2}^{c}>\varphi_{2}>\varphi_{2}^{c}
\end{equation}
\begin{equation}
\rho c^{2}-\sigma >0 \;\;\; for\; -\varphi_{2}^{c}<\varphi_{2}<\varphi_{2}^{c}
\end{equation}
at or near the throat.

At the angular horizons $\varphi_{2}=\varphi_{2}^{c}$
and $\varphi_{2}=2\pi-\varphi_{2}^{c}$, we have
$\rho c^{2}-\sigma =0$. All of these results have been obtained
for the specific metric where $\Phi=0$, but it is easy to check
that they are still valid for any metric component $g_{tt}$,
provided it is everywhere finite. It follows that for an observer
moving through the ringhole's throat with a sufficiently large
speed, $\gamma >> 1$, the energy density
\[\gamma^{2}(\rho c^{2}-\sigma)+\sigma\]
will be negative for $2\pi-\varphi_{2}^{c}>\varphi_{2}>\varphi_{2}^{c}$
and positive for $2\pi-\varphi_{2}^{c}>\varphi_{2}>\varphi_{2}^{c}$.

\vspace{.5cm}

\noindent (iv) From the above results one would expect lensing effects
to occur at or near the throat: the mouths would act like a diverging
lens for world lines along $2\pi-\varphi_{2}^{c}>\varphi_{2}>\varphi_{2}^{c}$,
and like a converging lens for world lines along
$-\varphi_{2}^{c}<\varphi_{2}<\varphi_{2}^{c}$. Of course, no lensing action
of the mouths could be expected at $\varphi_{2}=\varphi_{2}^{c}$
and $\varphi_{2}=2\pi-\varphi_{2}^{c}$.
Null-ray propagation reveals that focusing or defocusing of a bundle of
rays is governed by the integral of the stress-energy tensor
[3,8]. For the
mouths to defocus rays, such an integral must be negative; i.e.
$\int_{l_{1}}^{\infty}dle^{-\Phi}(\rho c^{2}-\sigma)<0$, for any
$l_{1}<0$, and positive if the mouths focus the rays. We shall
perform this integration at the extreme cases $\varphi=\pi$ and
$\varphi_{2}=0$. Using (3.11) and $\rho c^{2}-\sigma=
(\frac{n}{r})^{3}(T_{0}^{0}-T_{1}^{1})$, we first get for
$\varphi_{2}=\pi$ (for which $r=m=n=a+a$)
\[I_{\varphi_{2}=\pi}=\left.\frac{8\pi G}{c^{4}}\int_{l_{1}}^{\infty}dl(\rho c^{2}
-\sigma)\right|_{\varphi_{2}=\pi}=\]
\[\left[-\frac{l}{l^{2}+b_{0}^{2}}-\frac{b_{0}}{a^{2}}\arccos\left(\frac{b_{0}}{\sqrt{l^{2}+b_{0}^{2}}}\right)\right.\]
\begin{equation}
\left.\left.+\frac{l}{a^{2}}-\frac{1}{a}\ln\left(2l+2\sqrt{l^{2}+b_{0}^{2}}\right)\right]\right|_{l_{1}}^{\infty}.
\end{equation}
The condition $a\geq b>l$ implies that we must make $a\rightarrow\infty$
while performing the integration limits, so the above expression
reduces to $I_{\varphi_{2}=\pi}=-\frac{|l_{1}|}{l_{1}^{2}+b_{0}^{2}}<0$.
At $\varphi_{2}=0$ (for which $r=m=-n=a-b$), we likewise obtain
$I_{\varphi_{2}=0}=-I_{\varphi_{2}=\pi}>0$. Since at
$\varphi_{2}=\varphi_{2}^{c}$ and $\varphi_{2}=2\pi-\varphi_{2}^{c}$,
$n=0$ and hence $I_{\varphi_{2}=\varphi_{2}^{c}}
=0$, the above results can readily be generalized to:
\begin{equation}
\int_{l_{1}}^{\infty}dle^{-\Phi}(\rho c^{2}-\sigma)<0 \;\; for\; 2\pi-\varphi_{2}^{c}>\varphi_{2}>\varphi_{2}^{c}
\end{equation}
\begin{equation}
\int_{l_{1}}^{\infty}dle^{-\Phi}(\rho c^{2}-\sigma)>0 \;\; for\; -\varphi_{2}^{c}<\varphi_{2}<\varphi_{2}^{c}
\end{equation}
for any $\Phi$ which is everywhere finite.

Thus, as it was expected, the mouths act like a diverging lens for
$2\pi-\varphi_{2}^{c}>\varphi_{2}>\varphi_{2}^{c}$, and hence a
bundle of light rays that enters one mouth originally converging
to the direction tangent to the torus at points on
circunference at any
$2\pi-\varphi_{2}^{c}>\varphi_{2}>\varphi_{2}^{c}$
must emerge from the same points of the other ringhole's mouth
diverging from the same direction. Quite remarkably, unlike in
spherical wormholes, the ringhole's mouths act also like a
converging lens. This happens for
$-\varphi_{2}^{c}<\varphi_{2}<\varphi_{2}^{c}$, where a bundle
of light rays entering one mouth originally diverging emerges
from the other mouth converging into the given direction.

\section{\bf Ringhole time machine}
\setcounter{equation}{0}

The traversible ringhole space that we have considered in Sec. III
can be regarded as a Misner space [9,10] with the identified flat
planes replaced for identified tori when we give these tori
vanishing relative velocity $v=0$; i.e. it is equivalent to
extract two tori with given geometric parameters $a, b$ from
three-dimensional Euclidean space, and identify the torus
surfaces, so when you enter the surface of the right torus,
you find yourself emerging from the surface of the left torus,
and vice versa. In Minkowski spacetime, the ringhole is then
obtained identifying the two world conccentric tube pairs swept
out by the two tori, with events at the same Lorentz time
identified.

In order to convert a ringhole into time machine, one must set
one of the ringhole toroidal mouths in motion at a given speed
toward the other mouth, and identify then the two ringhole mouths
to each other. Before discussing the effects that the toroidal
shape of the ringhole mouths may cause in the pathologies of Misner
and spherical wormhole spaces, let us derive the spacetime metric
of such an accelerating ringhole. Assuming the right mouth to be the
moving mouth, just outside the right asymptotic
rest frame, the transformation of the ringhole coordinates into
external, Lorentz coordinates with metric $ds^{2}\simeq -dT^{2}
+dX^{2}+dY^{2}+dZ^{2}$, can be written as
\[T=T_{R}+v\gamma l\sin\varphi_{2}, \;\; Z=Z_{R}+\gamma l\sin\varphi_{2}\]
\[X=m_{l}\sin\varphi_{1},\;\; Y=m_{l}\cos\varphi_{1} ,\]
where $v=\frac{dZ_{R}}{dT_{R}}$ is the velocity of the right mouth;
$Z=Z_{R}(t), T=T_{R}(t)$ is the world line of the right mouth's
center with $dt^{2}=dT_{R}^{2}-dZ_{R}^{2}$, and $\gamma$ is the
relativistic factor $\gamma=(1-v^{2})^{-\frac{1}{2}}$. On the
other hand, just outside the left asymptotic rest frame, the
transformation is
\[T=t,\;\; Z=Z_{L}+l\sin\varphi_{2},\]
with the expressions for $X$ and $Y$ the same as just for outside the
right asymptotic rest frame. Here, $Z_{L}$ is the time-independent
$Z$ location of the left mouth's center. Then the metric inside
the accelerating ringhole and outside but near its mouths becomes
\begin{equation}
ds^{2}=-\left[1+glF(l)\sin\varphi_{2}\right]^{2}e^{2\Phi}dt^{2}
+dl^{2}+m_{l}^{2}d\varphi_{1}^{2}+b^{2}d\varphi_{2}^{2},
\end{equation}
where $g=\gamma^{2}\frac{dv}{dt}$ is the acceleration of the
right mouth and $\Phi$ is the same function as for the original
static ringhole. $F(l)$ is a form factor that vanishes in the
left half of the ringhole, $l\leq 0$, and rises monotonously
from 0 to 1 as one moves rightward from the throat to the
right mouth [3]. In obtaining (4.1) we have also used
$dv=g\frac{dt}{\gamma^{2}}$, $dt=\frac{dT_{R}}{\gamma}$ and
$d\gamma=vg\gamma dt$. This metric is a generalization to toroidal
symmetry of the metric used by Morris et al. for accelerating
spherical wormholes [3]. Since the symmetry axis of the ringhole
is taken again to be the axis $z$,
reduction from (4.1) to the spherical wormhole
metric can also be achieved by using the transformation $a\rightarrow 0$,
$\varphi_{2}\rightarrow\theta+\frac{\pi}{2}$, $\varphi_{1}
\rightarrow\phi$, as one should expect.

CTCs in accelerating ringhole space are originated at sufficiently
late times by exactly the same causes as in Misner or accelerating
spherical wormhole spaces [3,8]: on the left mouth the Lorentz
time, $T$, and the proper time, $\tau$, coincide, $\tau=T$, but
on the right mouth $\tau=\frac{T}{\gamma}$; this time dilation
must ultimately lead to CTCs which would only appear after reaching
a chronology (Cauchy) horizon. Such a horizon divides the full
ringhole spacetime into a chronal region and a non-chronal region
where CTCs appear. Like in Misner and accelerating spherical
wormhole spaces, there will be two families of timelike geodesics
in the chronal region: rightward geodesics and leftward geodesics,
both possessing their own chronology horizons and non-chronal
regions [11].

The mouth's lensing actions of the accelerating ringhole drastically
change the geometry of the chronology horizon for any $\varphi_{2}$,
except $\varphi_{2}=\varphi_{2}^{c}$ and $\varphi_{2}=2\pi-\varphi_{2}^{c}$,
at which cases the horizons
remain flat, null surfaces, like in Misner space.
In all the cases, one of the
Cauchy horizons is no longer a flat, null surface for any
$\varphi_{2}\neq\varphi_{2}^{c}$, $2\pi-\varphi_{2}^{c}$.
All that remains then is a
compact fountain [5,9] and, roughly speaking, a light cone at
one of the ringhole's mouths [12]. If you go in through the
ringhole along a world line where $2\pi-\varphi_{2}^{c}>\varphi_{2}
>\varphi_{2}^{c}$, the rightward chronology horizon and non-chronal
region are destroyed, so then the accelerating ringhole appears to be
only endowed with a leftward chronology horizon and non-chronal region.
In this case, there will be, roughly speaking, a future light cone
of point $P$ on the ringhole's left mouth (Fig. 3). The chronology
horizon becomes then the boundary of the future Cauchy development
of the fountain, $C_{R}$, generated by null geodesics which are
past-directed, to asymptote and enter the compact fountain $C_{R}$.
It can be said to be a compactly-generated horizon which is the
location of onset of rightward CTCs [5]. Thus, for world lines
with $2\pi-\varphi_{2}^{c}>\varphi_{2}
>\varphi_{2}^{c}$, everything happens like in accelerating
spherical wormholes and, therefore, if you could use an accelerating
ringhole to time travel along one such world lines, then you
would find unphysical violation of the classical averaged null
energy condition.

However, if you go in through the ringhole along a world line
which passes by the region
where $-\varphi_{2}^{c}<\varphi_{2}
<\varphi_{2}^{c}$ at or near the throat,
then the throat would act like a converging lens
for rightward propagation, and like a diverging lens for
leftward propagation (which takes place backward in time due
to time dilation of the accelerating right mouth), both with
focal length $f\simeq\frac{b_{0}}{2}$. The classical
instability caused by the pileup of leftward-propagating
waves on the leftward horizon can then be eliminated by
particle-defocusing if large separations between mouths
are used, so that $\frac{Bb_{0}}{2D}<<1$, where $B$ is the
Doppler frequency-shifting factor and $D$ is the distance between
ringhole mouths as measured along the fountain [3]. Therefore, the
leftward chronology horizon and non-chronal region are
destroyed and the accelerating ringhole becomes in this
case endowed with only the rightward chronology horizon and
non-chronal region.

For $-\varphi_{2}^{c}<\varphi_{2}
<\varphi_{2}^{c}$ at or near the throat,
the radical changes of the horizon geometry
are then different of those occurring in the case of world
lines passing always through the region
where $2\pi-\varphi_{2}^{c}>\varphi_{2}
>\varphi_{2}^{c}$ (see Sec. V). Now, roughly speaking, we have
a past light cone of point $Q$ also on the ringhole's right mouth
(Fig. 3). This past light cone lies in the past Cauchy development
of the compact fountain, $C_{L}$, and has as boundary a chronology
horizon compactly-generated by null geodesics which now are
future-directed, and asymptote and enter the fountain $C_{L}$.
Such a horizon is the location of onset of the chronal curves.
All the CTCS are situated in the future of this Cauchy horizon.

The crucial difference with respect to Misner and spherical
wormhole spaces is that in ringhole space, on world lines
with $-\varphi_{2}^{c}<\varphi_{2}
<\varphi_{2}^{c}$ at or near the throat,
you can traverse CTCs without ever encountering
a region with negative energy density.
At or in the close neighborhood of the angular horizons, the
energy density is always nonnegative and, since the ringhole
spacetime is asymptotically flat, the focusing theorem [13]
should then hold:
\[\frac{d^{2}A_{K}^{\frac{1}{2}}(P)}{d\lambda_{K}^{2}}\leq 0,\]
where $A_{K}$ is the cross-sectional area generated by the
intersection of the infinitesimal bundle $K$ of null-geodesic
generators of the angular horizons with a spacelike slice at point
$P$, and $\lambda_{K}$ is affine parameter along the bundle $K$.
Furthermore, it follows from (3.11) that for $b>>b_{0}$ the
energy density turns out to become positive for
$2\pi-\varphi_{2}^{c}>\varphi_{2}>\varphi_{2}^{c}$, and negative
for $-\varphi_{2}^{c}<\varphi_{2}<\varphi_{2}^{c}$, for
sufficiently large $\gamma$. Then, after
$b\simeq\sqrt{2}b_{0}$, the situation is inverted, the energy
density of region $2\pi-\varphi_{2}^{c}>\varphi_{2}>\varphi_{2}^{c}$
becoming negative and that of region
$-\varphi_{2}^{c}<\varphi_{2}<\varphi_{2}^{c}$ becoming positive,
at or near the throat, for large $\gamma$.

As one traverses the ringhole from one mouth to another
(Fig. 4), if we restrict to the region before
reaching the throat,
the angular horizons do not have any curvature
singularity which they can hit in the future and therefore
\begin{equation}
\frac{dA_{K}^{\frac{1}{2}}(P)}{d\lambda_{K}}\geq 0.
\end{equation}
So, like for event horizon of black holes, the total
cross-sectional area $A=\sum_{K}A_{K}$ of the ringhole angular
horizon halves before the throat
cannot decrease toward the future.

Any object with positive energy moving toward the throat through
the region $2\pi-\varphi_{2}^{c}>\varphi_{2}>\varphi_{2}^{c}$
(before reaching the cross-section at $b\simeq\sqrt{2}b_{0}$)
will be increasingly pulled gravitationally out from that
region, and then into the region
$-\varphi_{2}^{c}<\varphi_{2}<\varphi_{2}^{c}$, across the
angular horizons at around $\sqrt{2}b_{0}$, as the object
approaches the throat (Fig. 4).
One must choose then
the region where $2\pi-\varphi_{2}^{c}>\varphi_{2}>\varphi_{2}^{c}$,
up to $b\simeq\sqrt{2}b_{0}$, before reaching the throat,
as the causal past of the future null infinity, $\jmath^{-}(I^{+})$;
that is the location of observers inside the ringhole
that can send curves to future
null infinity $I^{+}$. Then, the union of angular horizons
at around $b\simeq\sqrt{2}b_{0}$,
before reaching the throat, will
be the future boundary $j^{-}(I^{+})$ of $\jmath^{-}(I^{+})$
for such observers.
The null generators of the union of the angular
horizons up to $b\simeq\sqrt{2}b_{0}$
can then be divided in infinitesimal bundles $K$ whose
intersection with spacelike slices defines the cross-sectional
areas $A_{K}$ satisfying (4.2),
so as closed two-surfaces which are the angular
horizons at the given times.
For observers in $\jmath^{-}(I^{+})$, the past singularity that
must occur
as $b\rightarrow a$ in region
$-\varphi_{2}^{c}<\varphi_{2}<\varphi_{2}^{c}$
does not matter for the stability of the horizons,
because bundles of null-geodesic generators of
the angular horizons can always be inserted in the future of
the past null infinity.

However, for observers traveling through the ringhole,
in the region $-\varphi_{2}^{c}<\varphi_{2}<\varphi_{2}^{c}$,
after the throat, there must also be
a curvature singularity in the future
infinity at $b\rightarrow a$ which the angular horizons will
unavoidably hit. From the focusing theorem it follows then [9,13]
\begin{equation}
\frac{dA_{K}^{\frac{1}{2}}(P)}{d\lambda_{K}}\leq 0.
\end{equation}

Although the singularity that lies in the future infinity of region
$-\varphi_{2}^{c}<\varphi_{2}<\varphi_{2}^{c}$ destroys
the horizons, such a
singularity would not be visible from the future null
infinity $I^{+}$ of the region
$2\pi-\varphi_{2}^{c}>\varphi_{2}>\varphi_{2}^{c}$, as it
is on the same asymptotic spacelike slice.
After reaching the throat, the object with positive energy
will be gravitationally pulled out from region
$-\varphi_{2}^{c}<\varphi_{2}<\varphi_{2}^{c}$, and into the
region $2\pi-\varphi_{2}^{c}>\varphi_{2}>\varphi_{2}^{c}$ again.
From $b\simeq\sqrt{2}b_{0}$
onward, after the throat, one must choose then the region
where $2\pi-\varphi_{2}^{c}>\varphi_{2}>\varphi_{2}^{c}$ as the causal
future of the past null infinity, $\jmath^{+}(I^{-})$; that is the
location of observers inside the ringhole
that can receive curves from the past null
infinity $I^{-}$. In this case, for such observers
the union of angular horizons
at around $b\simeq\sqrt{2}b_{0}$, after the throat, will
be the past boundary $j^{+}(I^{-})$ of $\jmath^{+}(I^{-})$, and, thereafter,
their infinitesimal bundles of null-geodesic generators will, by
intersecting spacelike slices, define cross-sectional areas
$A_{k}$ satisfying (4.3), and closed two-surfaces that are
the horizons at the given times.

We can then conclude that
gravitational interaction inside the ringhole sets
the kind of itineraries that can be followed when traversing
a ringhole, without ever encountering a region with "exotic"
matter (Fig. 4).
If you go in through the
ringhole moving mouth and come out through the stationary mouth
by traversing a CTC along one such itineraries, you will go
backward in time, without encountering yourself surrounded
by matter that violates the averaged weak energy condition, or finding
any classical instability.
You can do similarly safe classical
travels into the future if you go in through the stationary mouth
and exit through the mouth that has moved, along these world lines.

\section{\bf Vacuum fluctuations}
\setcounter{equation}{0}

Let us consider the effects of
electromagnetic quantum vacuum polarization
in accelerating ringholes. For this case, the point-splitting
regularized Hadamard two-point function for a quantized, massless,
conformally-coupled scalar field in the spacetime depicted in
Fig. 3, for regions where the curvature vanishes [7], can be written
in the form
\[G_{reg}^{(1)\pm}(x,x')=\]
\begin{equation}
\sum_{N=1}^{\infty}\frac{\xi}{4\pi^{2}D}\left(\frac{b\xi}{2D}\right)^{N-1}
\left(\frac{1}{\lambda_{N}^{\pm}(x,x')}+\frac{1}{\lambda_{N}^{\pm}(x',x)}\right),
\end{equation}
where
\[\xi=\left(\frac{1-v}{1+v}\right)^{\frac{1}{2}}<1,\]
$D$ is the spatial length of a geodesic that connects points $x$ and
$x'$ by traversing the ringhole, and
\[\lambda_{N}^{\pm}(x,x')=\xi^{N}\frac{\sigma_{N}^{\pm}}{\zeta_{N}}\]
in which $\zeta_{N}=D\left(\frac{1-\xi^{N}}{1-\xi}\right)$ and
$\sigma_{N}^{\pm}$ is the $N$th geodetic interval between $x$ and
$x'$ for (+) $2\pi-\varphi_{2}^{c}>\varphi_{2}>\varphi_{2}^{c}$ and
(-) $-\varphi_{2}^{c}<\varphi_{2}<\varphi_{2}^{c}$. $\lambda_{N}^{\pm}$
has been evaluated by the method of Hiscock and Konkowski used by
Kim and Thorne [7]. In our case, by using Fig. 1, where the symmetry
axis of the mouths is now taken to be
$Y$, and the covering space [8] for the $Y-T$
plane of the ringhole spacetime, we can calculate the displacements
at fixed times $T'$ and $T$ of, respectively, copy 0 of $x'$ and
copy $N$ of $x$ from the covering-space throat location for
$\varphi_{1}=0$, i.e.
\[\bigtriangleup\tilde{Y}_{0}^{\pm}(x')=-(a\pm b)+m\]
\[\bigtriangleup\tilde{Y}_{N}^{\pm}(x)=\xi^{-N}\left[(a\pm b)-m\right],\]
and hence the corresponding geodetic intervals,
\[\sigma_{N}^{\pm}=\zeta_{N}\left[\left(\bigtriangleup\tilde{Y}_{N}^{\pm}\xi^{N}
-T\right)\xi^{-N}-\left(\bigtriangleup\tilde{Y}_{0}^{\pm}-T'\right)\right],\]
when the points $x'$ and $x$ are not on the symmetry axis, become
\[\sigma_{N}^{\pm}(x,x')=\]
\begin{equation}
\zeta_{N}\left\{\xi^{-N}\left[\pm b(1\pm\cos\varphi_{2})-T\right]
+\left[\pm b(1\pm\cos\varphi_{2})+T'\right]\right\}.
\end{equation}
If the points $x'$ and $x$ are also slightly off the throat in the
$Y$-direction, we finally have
\[\lambda_{N}^{\pm}(x,x')=\]
\begin{equation}
\pm 2b(1\pm\cos\varphi_{2})+Y-T-(Y'-T')\xi^{N},
\end{equation}
where again the upper sign (+) stands for
$2\pi-\varphi_{2}^{c}>\varphi_{2}>\varphi_{2}^{c}$, and the lower
sign (-) does for
$-\varphi_{2}^{c}<\varphi_{2}<\varphi_{2}^{c}$, with $T=\tau$, $Y$
the Lorenzt coordinates of two-dimensional flat, Minkowski
spacetime of the world line near the left mouth, and
$T'-\gamma(T-D)+v\gamma(Y-D)$, $Y'=\gamma(Y-D)+v\gamma(T-D)$ the
equivalent coordinates on a world line near the right mouth which
is at rest at the origin.

According to Kim and Thorne [7], a $N$th-polarized hypersurface,
$H_{N}$, is formed by those events that join to themselves through
closed null geodesics by traversing the ringhole $N$ times. On
these hypersurfaces the quantum polarization of vacuum should
diverge [7,12]. Then, upon collapsing the points $x'$ and $x$
together, it turns out that there will be polarized hypersurfaces
at times given by $\sigma_{N}^{\pm}=0$; i.e.:
\begin{equation}
T_{H_{N}}^{\pm}=\pm\left(\frac{1+\xi^{N}}{1-\xi^{N}}\right)b(1\pm\cos\varphi_{2}).
\end{equation}

The chronology (Cauchy) horizons, $H^{\pm}$, depicted in Fig. 3,
are the limit as $N\rightarrow\infty$ of the times $T_{H_{N}}^{\pm}$,
and respectively nest the corresponding polarized hypersurfaces
defined at times given by (5.4). On the symmetry axis, i.e. for
$\varphi_{2}=\pi$ and $\varphi_{2}=0$, all polarized hypersurfaces
$H_{N}$ occur at the same time only at $T=0$; away from the
symmetry axis, one meets the $H_{N}$ one after another beginning
with arbitrarily large $N$ and ending with $N=1$ as $T$ increases
if $2\pi-\varphi_{2}^{c}>\varphi_{2}>\varphi_{2}^{c}$, or as $T$
decreases if $-\varphi_{2}^{c}<\varphi_{2}<\varphi_{2}^{c}$. The
nesting of the $H_{N}$'s in the chronology horizon $H^{+}$ occurring
at time
\[T_{H^{+}}=+b(1+\cos\varphi_{2})\]
for $2\pi-\varphi_{2}^{c}>\varphi_{2}>\varphi_{2}^{c}$ guarantees
that, an observer entering the region of CTCs will pass first
through the chronology horizon $H^{+}$, and then successively
through various $H_{N}$'s. In the case
$-\varphi_{2}^{c}<\varphi_{2}<\varphi_{2}^{c}$, the $H_{N}$'s
are located in the chronology horizon $H^{-}$ occurring at time
\[T_{H^{-}}=-b(1-\cos\varphi_{2}),\]
so that the observer will first pass through successive $H_{N}$'s
and then enters the chronology horizon $H^{-}$. Perhaps in this
case the observer could avoid experiencing the strong peaks of
vacuum polarization by carefully choosing the moment at which
she starts moving along a world line leading to CTCs.

Anyway, we see that semiclassical instabilities would also
appear in the accelerating ringhole in both regions separately.
Whether or not quantum-gravity effects or angular-horizon
effects can cut off these divergences independently on each
side of the angular horizon is a rather open question that has been
the subject of much debate in recent years [5,7].

We note
nevertheless that the renormalized stress-energy tensor
$T_{\mu\nu}^{\pm}$, obtained by differentiating twice
$G_{reg}^{(1)\pm}$ [7], on polarized hypersurfaces
with the same $N$ should satisfy
$T_{\mu\nu}^{+}=-T_{\mu\nu}^{-}$, when the angle $\varphi_{2}$
in the interval $2\pi-\varphi_{2}^{c}>\varphi_{2}>\varphi_{2}^{c}$
can be obtained from the corresponding angle in the interval
$-\varphi_{2}^{c}<\varphi_{2}<\varphi_{2}^{c}$ by summing $\pi$
to the latter angle. From (5.4) we obtain in this case
that the difference
between the occurrence times of polarized hypersurfaces for the
same $N$, nested in the two approximate light cones of Fig. 3
is given by
\[\bigtriangleup T_{H_{N}}=T_{H_{N}}^{+}-T_{H_{N}}^{-}
=2b\left(\frac{1+\xi^{N}}{1-\xi^{N}}\right)(1+\cos\varphi_{2}),\]
where $2\pi-\varphi_{2}^{c}>\varphi_{2}>\varphi_{2}^{c}$.
This precisely is the time the particles on the $N$th-polarized
hypersurface at a given $\varphi_{2}^{(i)}$
on the side of the angular horizon where
$-\varphi_{2}^{c}<\varphi_{2}<\varphi_{2}^{c}$ would take to
go from their original location to that of the $N$th-polarized
hypersurface at $\varphi_{2}=\varphi_{2}^{(i)}+\pi$, on the other side
of the angular horizon, reaching the latter hypersurface at
exactly the time it is created.

It follows from the discussion in Sec. IV that,
at or near the throat, everything having positive energy density,
originally in the region
$2\pi-\varphi_{2}^{c}>\varphi_{2}>\varphi_{2}^{c}$,
should cross the angular
horizons into the region $-\varphi_{2}^{c}<\varphi_{2}<\varphi_{2}^{c}$,
and everything
having negative energy density, originally in the region where
$-\varphi_{2}^{c}<\varphi_{2}<\varphi_{2}^{c}$, should cross the
horizons in the opposite sense to get into the region
$2\pi-\varphi_{2}^{c}>\varphi_{2}>\varphi_{2}^{c}$.
Moreover, since polarized hypersurfaces in region
$2\pi-\varphi_{2}^{c}>\varphi_{2}>\varphi_{2}^{c}$ occur
later than those in region
$-\varphi_{2}^{c}<\varphi_{2}<\varphi_{2}^{c}$,
there will be then a flux of vacuum-polarized particles created
by electromagnetic quantum fluctuations through the horizons,
at or near the throat: that from the region where
$-\varphi_{2}^{c}<\varphi_{2}<\varphi_{2}^{c}$ to the region
where $2\pi-\varphi_{2}^{c}>\varphi_{2}>\varphi_{2}^{c}$. Just
once these particles have traveled an angular path
$\bigtriangleup\varphi_{2}=\pi$ during a time $\bigtriangleup T_{H_{N}}$,
crossing the angular horizons at $\varphi_{2}\simeq\frac{\pi}{2}$,
they would reach the location where the otherwise identical
vacuum-polarized particles with the opposite-sign energy are
being created on polarized hypersurfaces with the same $N$, at
exactly the time their counterparts arrive, so that, in
the limit $\frac{b_{0}}{a}\rightarrow 0$, the quantum-vacuum
polarizations occurring on both sides of the angular horizons
would exactly cancel each other, leaving finally a vanishing
polarized stress-energy tensor.

\section{\bf Conclusions}

In this paper we have explored the possibility that wormholes
with the topology of a torus may exist and be used to construct
a time machine. We call these wormholes ringholes. Ringholes
with extremely large spacetime curvature could possibly be also
spontaneously created in the quantum-gravity regime, and then
pulled out from the quantum spacetime foam by a future technological
process that would make them grow up to classical sizes [3]. The
probability for the existence of such constructs in the foam is
not known, but it has been recently established both from a
theorem of Gannon [14] and by numerical computation [15] that,
in the collapse of rotating collisionless matter, a toroidal
horizon can be formed before the usual spherical horizon, provided
we respect asymptotic flatness and the dominant energy condition.

By implicitly assuming the above or other possible mechanisms for
ringhole creation, in this paper we have first considered the
spacetime static metric corresponding to a ditribution of matter
with the topology of a torus whose surface area and shape are
kept constant. This solution shows two angular horizons on
circunferences which are in between the maximum and minimum
circunferences of the torus and parallel to them.

We then considered the static metric of traversible ringholes.
These tunnels possess two distinct regions, separated by the
angular horizons, which show different behaviours. On the
region surrounding the maximum circunference,
near the throat the ringhole
acts like a diverging lens and is associated with an "exotic"
stress-energy tensor having negative energy density. On the
region surrounding the minimum circunference
near the throat, the ringhole
acts like a converging lens and associates with an ordinary
stress-energy tensor allowing for positive energy density only.

Upon converting a ringhole into time machine by making one
ringhole mouth to move with respect to the other and
identifying the two mouths, it has been shown that there exists
a chronology horizon on the inner region which opens up to a
non-chronal region where time travels through closed timelike
curves can be performed without ever encountering any violation
of the averaged null energy condition. One could regard the
angular horizons as implying some sort of {\it censorship}
protecting the observers in the vecinity of the minimum
circunference of the torus against experiencing the action
of matter with negative energy density of the other angular
horizon side, at or near the ringhole throat.

Polarized hypersurfaces with diverging stress-energy tensor have
been shown to exist on both sides of the angular horizons:
vacuum-polarized modes with positive energy are all located on
the side of the maximum circunference, and those with negative
energy only are on the other side of the angular horizons.
Near the throat, the horizons have been found to bahave
as one-directional membranes, so that all these
modes would be expected to eventually end up all together confined
in the region surrounding the maximum toroidal
circunference, where they would
exactly cancel each other in the limit of
large tori with small throats, so that, in this case,
the quantum vacuum polarization
became unable to prevent ringholes and hence closed timelike
curves to exist. Whether or not time travels through these
ringholes would then entail an irretrievable loss of coherence [16]
making the travels useless and very dangerous is a matter to
be investigated.

\acknowledgements

\noindent For helpful comments, the author thanks C. Sigenza
and G. Mena Marug n. This
research was supported by DGICYT under research projects N§
PB94-0107 and N§ PB93-0139.


\pagebreak

\begin{center}
{\bf Legend for Figures}
\end{center}

\vspace{.5cm}

\noindent $\bullet$ Fig. 1. Cartesian coordinates on the
two-dimensional torus. Any point $P$ on the torus surface can be
labeled by parameters $a, b, \varphi_{1}, \varphi_{2}$.

\vspace{.5cm}

\noindent $\bullet$ Fig. 2. Embedding diagram for a spacetime
ringhole that connects two regions of the universe.

\vspace{.5cm}

\noindent $\bullet$ Fig. 3. A spacetime ringhole with the right mouth
moving toward the left. Also depicted are the fountains ($C$) and
the chronology horizons ($H^{\pm}$) for the regions on both
sides of the angular horizons at $\varphi_{2}=\varphi_{2}^{c}$
and $\varphi_{2}=2\pi-\varphi_{2}^{c}$.

\vspace{.5cm}

\noindent $\bullet$ Fig. 4. Half of a $\varphi_{1}$ = Const.
section of a ringhole depicting the different regions inside
it (see the text). The dashed line is an example of a trajectory
that never passes through a region with negative energy density,
even for large values of the relativistic factor $\gamma$.


\begin{references}
\bibitem {11} This can be rather shown if we take Minkowski spacetime
as the covering space [8,9] for ringhole space. This can be constructed
by lining up a sequence of copies of ringhole space, side by side, each one
boosted by speed $v$ relative to the last one.
\end{references}
\end{document}